\documentclass[fleqn,usenatbib]{mnras}

\usepackage[T1]{fontenc}
\usepackage{ae,aecompl}

\usepackage{graphicx}	
\usepackage{amsmath}	
\usepackage{amssymb}	
\usepackage{newtxtext,newtxmath}
\usepackage{xcolor}

\newcommand{\chandra}{\textit{Chandra}}
\newcommand{\nustar}{\textit{NuSTAR}}
\newcommand{\suzaku}{{\it Suzaku}}

\newcommand{\xmm}{{\it XMM-Newton}}

\definecolor{rednew}{RGB}{243,60,83}
\definecolor{redchunk}{RGB}{202,52,51}
\newcommand{\redch}{\textcolor{black}}
\newcommand{\red}{\textcolor{black}}
\newcommand{\ergs}{erg~cm$^{-2}$~s$^{-1}$}

\newcommand{\src}{MCG$-$06}
\newcommand{\srcfull}{MCG$-$06-30-15}

\newcommand{\second}{\textcolor{black}}

    \title[Spectral Fitting of \srcfull]{Black Hole Spin Measurements Based on a Thin Disc Model with Finite Thickness I. An example study of \srcfull}

\author[J. Jiang et al.]{
Jiachen Jiang,$^{1}$\thanks{E-mail: jj447@cam.ac.uk} Askar B. Abdikamalov$^{2,3,4}$, Cosimo Bambi$^2$ 
\newauthor and Christopher S. Reynolds${^1}$ 
\\
$^{1}$Institute of Astronomy, University of Cambridge, Madingley Road, Cambridge CB3 0HA, UK\\
$^{2}$Department of Physics, Fudan University, 2005 Songhu Road, Shanghai 200438, China\\
$^{3}$Ulugh Beg Astronomical Institute, Tashkent 100052, Uzbekistan\\
$^{4}$Institute of Fundamental and Applied Research, National Research University TIIAME, Kori Niyoziy 39, Tashkent 100000, Uzbekistan
}

\date{Accepted XXX. Received YYY; in original form ZZZ}

\pubyear{2021}

\begin{document}
\label{firstpage}
\pagerange{\pageref{firstpage}--\pageref{lastpage}}
\maketitle

\begin{abstract}
We present a re-analysis of the \xmm\ and \nustar\ observing campaign for the well-studied, X-ray-bright AGN \srcfull. In particular, we consider a disc model with finite thickness. By fitting the disc reflection spectra in the data, we obtain a black hole spin of 0.87--0.99 (90\% confidence range) after taking the thickness of the disc into consideration. Spectral models with a grid of mass accretion rate from $0$ to $30\%\dot{M}_{\rm Edd}$ are calculated for \srcfull. This result is obtained by considering a free disc reflection fraction parameter $f_{\rm refl}$ and is consistent with previous measurements based on razor-thin disc models. Besides, an isotropic, point-like geometry, i.e. the `lamppost' geometry, is assumed for the corona in our model. We find that such a geometry overestimates $f_{\rm refl}$ in the data. Therefore, thin disc models with consistent `lamppost' values of $f_{\rm refl}$ provide a worse fit than ones with a free $f_{\rm refl}$ parameter. We discuss possible reasons for the discrepancy between the observed and theoretical values of $f_{\rm refl}$ at the end of the paper. Modifications for the over-simplified lamppost model might be needed when the thickness of the thin disc is considered in future work.
\end{abstract}




\section{Introduction}

\subsection{Disc Reflection Spectroscopy and Black Hole Spin}

The X-ray continuum emission of Seyfert active galactic nuclei (AGN)  shows a power-law shape, which originates in the up-scattering Comptonisation processes of lower energy disc photons in the black hole (BH) corona \citep[e.g.][]{maisack93}. The optically thick disc is illuminated by the hot corona and produces a reprocessed spectrum on its surface, which is often referred to as the disc `reflection' spectrum. The most prominent features of the disc reflection spectrum are the Fe~K$\alpha$ emission line at 6.4\,keV \citep[e.g.][]{tanaka95} and the back-scattering Compton hump above 10\,keV \citep[e.g.][]{nandra90}. The Fe~K$\alpha$ emission line of the disc shows a distinctive shape rather than a narrow line, which results from the strong relativistic effects in the vicinity of a BH  \citep{fabian04}. Such a broad Fe~K$\alpha$ emission line has been seen in not only a handful of  AGN \citep[e.g.][]{larsson08, brenneman11, tan12, risaliti13, parker14, walton13, jiang18d} but also X-ray binaries \citep[e.g.][]{fabian89,cackett13,miller13}.

The modelling of disc reflection spectra involves many detailed scientific calculations. For instance, the properties of the accretion disc \citep[e.g.][]{garcia16,jiang18,jiang19b}, the geometry of the corona \citep[e.g.][]{wilkins14,gonzalez17} and the spacetime around a BH \citep[e.g.][]{berti15,bambi17} all have to be taken into consideration. The disc reflection model is applied accordingly to study different topics. 

One of the most frequent applications of disc reflection spectroscopy is the measurement of BH spin. The extent of the broad Fe~K$\alpha$ emission line to low energies, the `red wing', is determined by how close the line emission region of the disc is to the BH \citep{fabian89}. When the inner edge of the disc is closer to the BH, the disc Fe~K$\alpha$ emission shows a more extended red wing due to gravitational redshift. Assuming the inner edge of the disc is at the innermost stable circular orbit (ISCO), one can measure the spin of the BH by using disc reflection spectropy based on the simple correlation between BH spin and the radius of ISCO \citep[see latest reviews in][]{reynolds19,bambi21}.

\subsection{The Thin Disc Model with Finite Thickness}

Previous disc reflection models assumed a simple thin disc model with infinitely small height for simplification. There have been few attempts to fit the X-ray data of AGN using a model with finite disc thickness as expected in reality because of the difficulty of modelling more complex geometries.

In the standard thin disc model, the pressure scale height of a radiation pressure-dominated disc at $r$ is

\begin{equation}
    H=\frac{3}{2}\frac{\dot{m}}{\epsilon}[1-(\frac{r_{\rm ISCO}}{r\sin(\theta)})^{1/2}],
 \label{Hva}
\end{equation}

\redch{where $\dot{m}$ is mass accretion rate in units of Eddington accretion rate ($\dot{m}=\dot{M}/\dot{M}_{\rm Edd}$)}, $\epsilon$ is the radiative efficiency of the disc and $r\sin(\theta)$ is the pseudo-cylindrical radius \citep{shakura73}. Materials at the same $r\sin(\theta)$ share the same angular velocity. This solution is calculated assuming the flat Newtonian spacetime but very similar with the relativistic solution \citep[e.g.][]{novikov73,pariev98}. 

\begin{figure*}
    \centering
    \includegraphics[width=17cm]{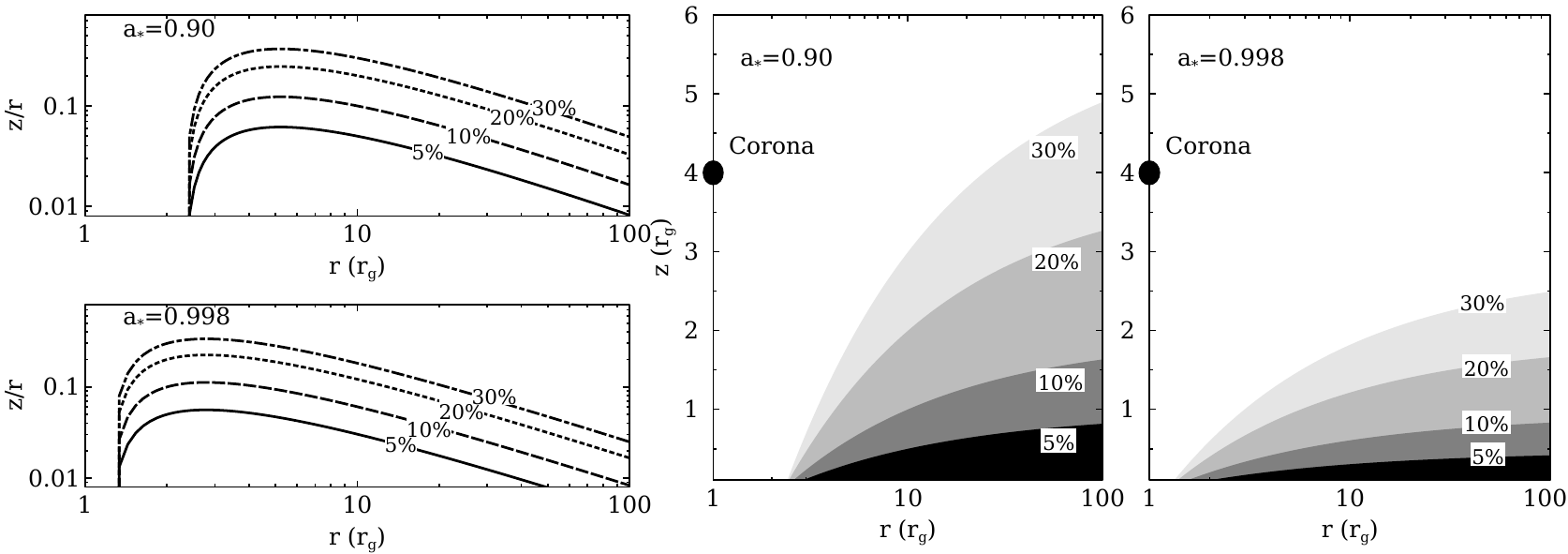}
    \caption{Disc shapes at different $\dot{m}$. The left two panels show the disc thickness profile $z/r$ for $a_{*}=0.90$ (top) and $a_{*}=0.998$ (bottom). The inner radius of the disc is assumed to be at ISCO. At higher $\dot{m}$, the disc is thicker. The right two panels show schematic images of the thin disc models for different $\dot{m}$ and $a_*$. The black circles show that the corona is located on the spinning axis of the BH in the lamppost geometry.}
    \label{pic_shape}
\end{figure*}

As in previous reflection model in \citet{pariev98,taylor18a}, we assume the X-ray reflection atmosphere of the disc is at $z=2H$. We show the profiles of $z/r$ for various $\dot{m}$ and $a_{*}$ in Fig.\,\ref{pic_shape}. At $\dot{m}=5\%$, $z/r$ reaches 4\% in the inner disc region and starts decreasing in the outer disc. In comparison, $z/r$ reaches a higher value of 30\% in the inner disc region, indicating a thicker disc at higher $\dot{m}$. When $\dot{m}$ is higher than 30\%, the disc may not hold the thin disc geometry as in Eq.\,\ref{Hva} \citep{shakura73}. We, therefore, only discuss the cases of $\dot{m}<30\%$ in this work.

In Fig.\,\ref{pic_shape}, we show that the inner radius of the disc, which is assumed to be at ISCO, is closer to the BH at the maximum BH spin than a modest BH spin. The disc is also thinner at a higher spin than a lower spin, suggesting that $\dot{m}$ has a smaller impact on disc thickness when the BH spin is high because of a higher radiative efficiency. For instance, the height of the disc is up to 5\,$r_{\rm g}$ within a radius of 100\,$r_{\rm g}$ for $a_{*}=0.9$ while the height reaches only 2.5\,$r_{\rm g}$ at most for $a_{*}=0.998$.

The thickness of the disc in the standard disc model \citep{shakura73} is comparable to the observed size of the coronal region in AGN. X-ray reverberation studies suggest that the corona has to be very compact within a region of approximately 10\,$r_{\rm g}$ \citep[e.g.][]{fabian09,demarco13,kara17,alston20}. So the thickness of the disc may play an important role in disc reflection modelling.

\begin{figure}
    \centering
    \includegraphics[width=\columnwidth]{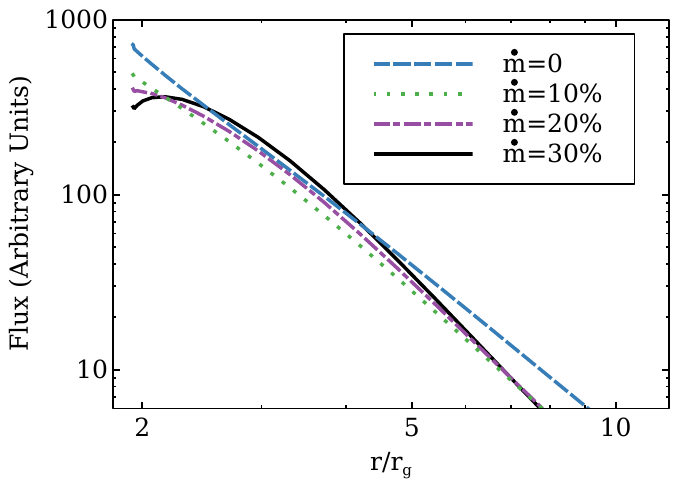}
    \caption{Disc emissivity profiles for different values of $\dot{m}$ when the thickness of the disc is considered. The corona is  at $h=4$\,$r_{\rm g}$ and the BH spin is $a_{*}=0.95$ in the calculations.}
    \label{pic_q}
\end{figure}

The effects of disc thickness on resulting reflection spectra were calculated in \citet{pariev98,wu07,tripathi21}. Numerical ray-tracing techniques were used. They calculated $g$-factors, \redch{the energy shift from the disc to the observer}, on the surface of the inner accretion disc accordingly. However, the disc emissivity profile was approximated as a power law. Further studies by \citet{taylor18a} show that the observed emissivity profile is also an essential indicator of disc thickness. For example, we compare the emissivity profiles of a razor-thin disc and ones with finite thickness in Fig.\,\ref{pic_q}. The emissivity profile of the disc is flatter than the one for a razor-thin disc as shown in Fig.\,\ref{pic_shape}. 

To calculate the emissivity profile of a thin disc, one needs to assume the coronal geometry first. We consider the same geometry as in \citet{taylor18a} by adapting the simplest and most understood `lamppost' geometry in our model, where the corona is a point source located on the spinning axis of the BH \citep{martocchia96}. \redch{In this set-up, the corona emits isotropic emission in its rest frame.} The lamppost geometry successfully explains the X-ray data of many AGN \citep[e.g.][]{miniutti04,martocchia02,niedzwiecki08} and is thought to be related to the base of the jet \citep[e.g.][]{ghisellini04} or pair productions in magnetosphere around the BH \citep[e.g.][]{hirotani98,chen20}. Although the lamppost model is used in this work, we will also investigate whether the lamppost geometry is sufficient to explain the data at the end of this paper.


\subsection{\srcfull}

To investigate the effects of disc thickness on the reflection modelling of AGN spectra, we start with the most studied X-ray bright AGN in \srcfull\ (\src\ hereafter). \src\ is a Seyfert 1.2 galaxy \citep{bennert06} at $z=0.008$ \citep{fisher95}. The BH mass of \src\ is estimated to be  $1.6\pm0.4\times10^{6}M_{\odot}$ based on optical reverberation measurements \citep{bentz16}. 


In the X-ray band, \src\ is the first AGN where broad Fe~K$\alpha$ emission was studied in detail \citep{tanaka95, iwasawa96, fabian02}, allowing measurements of its BH spin using disc reflection spectroscopy. Multiple studies indicate that the spin of the BH in \src\ is high \citep[e.g. $a_*$>0.90,][]{brenneman06,vaughan04,young05,reynolds05,miniutti07,marinucci14}. Detailed modelling suggests two components in the Fe~K emission of \src, one from the innermost accretion region close to the ISCO and one from a distant neutral reflector \citep{ballantyne03}. Soft X-ray reverberation lags are also discovered in \src\ \citep{demarco13,kara14}. They are related to the light travel time difference between the disc reflected light and the coronal X-ray continuum. \redch{However, no significant evidence of Fe~K reverberation lags has been found in \src\ \citep{kara14}.}

Previously, \citet{tripathi21} applied the thin disc model with finite thickness to the X-ray data of \src. A broken power law was used to model the emissivity profile of its disc. They achieved consistent BH spin measurements for \src\ as in previous work. However, the emissivity profile is also affected by the geometry of the disc as introduced above. We, therefore, reanalyse the data of \src\ and consider the same geometry set-up as in \citet{taylor18a}--a thin disc with finite thickness shouldered by the lamppost corona. 



In Section \ref{data}, we introduce our data reduction processes. In Section \ref{free}, we present a thin disc model with finite thickness for the spectral data of \src. The reflection fraction parameter is treated as a free parameter during the fit. In Section \ref{link}, we further discuss the effects of disc thickness on the reflection fraction parameter of the model. Models with consistent values of reflection fraction assuming the lamppost geometry are used to fit the data of \src. In Section \ref{discuss}, we discuss and conclude our results.

\section{Data Reduction} \label{data}

In this work, we consider the \xmm\ \citep{jansen01} and \nustar\ \citep{harrison13} observing campaign of \src\ in 2013. The same observations were analysed in \citet{marinucci14,tripathi21}. A full list of the observations is in Table\,\ref{tab_obs}.

The EPIC data are reduced using V19 of the \xmm\ Science Analysis System (SAS) software package. The version of the calibration files  is v.20201028. We first generate a clean event file by running EPPROC. Then, we select good time intervals by filtering out the intervals that are dominated by flaring particle background. These high-background intervals are where the single event (PATTERN=0) count rate in the >10~keV band is larger than 0.4 counts~s$^{-1}$ for pn data. By running the EVSELECT task, we select single and double event lists from a circular source region of 35 arcsec. Background spectra are extracted from a nearby circular region of 60 arcsec. No obvious evidence of pile-up effects has been found in pn data. We do not consider MOS data because they have significant pile-up effects. The pile-up effects are checked using the EPATPLOT tool in SAS. Last, we create redistribution matrix files and ancillary response files by running RMFGEN and ARFGEN. The spectra are grouped to have a minimum of 20 counts per bin and oversample by a factor of 3.

We reduced the \nustar\ data using the \nustar\ Data Analysis Software (NuSTARDAS) package and calibration data of V20211103. The energy spectra of \src\ were extracted for both the FPMA and FPMB detectors from a 100$''$ radius circle centered on the source, while the background spectra were extracted from a nearby circular region of the same size. We consider the 3--50\,keV band of the two FPM spectra. The FPM data above 50\,keV are dominated by the background.

In the soft X-ray band, \src\ is known to show complex absorption features from dust and warm absorbers \citep[e.g.][]{lee02, turner03,young05,chiang11}. Besides, the inclusion of soft X-ray data may introduce extra uncertainty to our spin measurements \citep[e.g. by modelling the soft excess emission, ][]{jiang19b}. Therefore, we focus on only the hard X-ray data of \src\ above 3\,keV, where the broad Fe~K$\alpha$ emission line and the Compton hump are. We show in Section\,\ref{free} that consistent measurements as in previous work can still be achieved when we consider only the hard X-ray data.

\begin{table*}
    \centering
    \begin{tabular}{cccccc}
    \hline\hline
    \nustar\ Obs ID & Date & Time (ks) & \xmm\ Obs ID & Date & Time (ks) \\
    \hline
    60001047002& 2013-01-29 & 23 & 0693781201 & 2013-01-29 & 134 \\
    60001047003& 2013-01-30 & 127 & 0693781301 & 2013-01-31 & 134\\
    60001047005& 2013-02-02 & 30 &0693781401 & 2013-02-02 & 49\\
    \hline\hline
    \end{tabular}
    \caption{\nustar\ and \xmm\ observations of \src\ analysed in this work. }
    \label{tab_obs}
\end{table*}

\section{Spectral Analaysis} \label{free}

We use XSPEC V.12 \citep{arnaud85} for spectral analysis and $\chi^{2}$ to estimate goodness of fit. We start our analysis with the simple razor-thin disc model. Then we consider a thin disc model with a grid of $\dot{m}$ at 5\%, 10\%, 20\% and 30\%. We will also compare our results with previous spin measurements.

\subsection{The Razor-Thin Disc Model} \label{m0}

We model the spectra of \src\ by following previous conclusions \citep[e.g.][]{young05, ballantyne03, marinucci14}: the Fe~K$\alpha$ emission of \src\ consists of two components, one from a distant neutral reflector and one from the innermost accretion disc. The former is calculated using the \texttt{xillver} model \citep{garcia10}. The ionisation parameter\footnote{The ionisation parameters are reported in units of erg cm s$^{-1}$ hereafter.} of this component is fixed at $\log(\xi$/erg cm s$^{-1})=0$. The reflection spectrum from the inner accretion disc is calculated using the \texttt{relxilllpth\_nk} \citep{ab20}, an extended version\footnote{\citet{tripathi21} applied the \texttt{relxillth\_nk} model to the data of \src. A more general metric than the Kerr metric was used, where an extra deformation parameter was considered. When the deformation parameter is zero, this metric is identical to the Kerr metric \citep{kerr63}. It has been found that the data of \src\ are consistent with the Kerr metric \citep{tripathi21}. So we only consider the Kerr metric in this work for simplicity.} of the \texttt{relxill\_nk} package \citep{bambi17,abdikamalov19}. To model a razor-thin disc, the $\dot{m}$ parameter of the model is set to be zero. Free parameters include the dimensionless spin of the BH ($a_*$), the inclination angle of the disc ($i$), the ionisation and the iron abundance of the disc ($\xi$ and $Z_{\rm Fe}$) and the height of the corona ($h$). \redch{The inner radius of the disc is set to be the ISCO.} The illuminating spectrum of \texttt{relxilllpth\_nk} is a power law ($\Gamma$) with an exponential high-energy cutoff at Ecut. The emissivity profile of the disc is calculated consistently according to the the geometry of the disc and the height of the corona. Note that the reflection fraction parameter $f_{\rm refl}$ is a free parameter in this section\footnote{The reflection fraction $f_{\rm refl}$ is defined as the ratio between the flux that reaches the disc and infinity \citep{dauser16}. We will consider consistent values of $f_{\rm refl}$ in Section \ref{link}.}. 

A \texttt{constant} model is used to account for cross-calibration uncertainties between instruments. The \texttt{tbabs} model is used to account for Galactic absorption. \redch{We consider the interstellar medium abundances calculated by \citet{wilms00} in the \texttt{tbabs} model.} The Galactic column density along the line of sight towards \src\ is estimated to be $4.7\times10^{20}$\,cm$^{-2}$ \citep{willingale13}. The effect of such a low column density is confined in the soft X-ray band, e.g. <2\,keV. Since we do not include soft X-ray data, we therefore fix $N_{\rm H}$ at this value in the following analysis.

\begin{figure}
    \centering
    \includegraphics[width=\columnwidth]{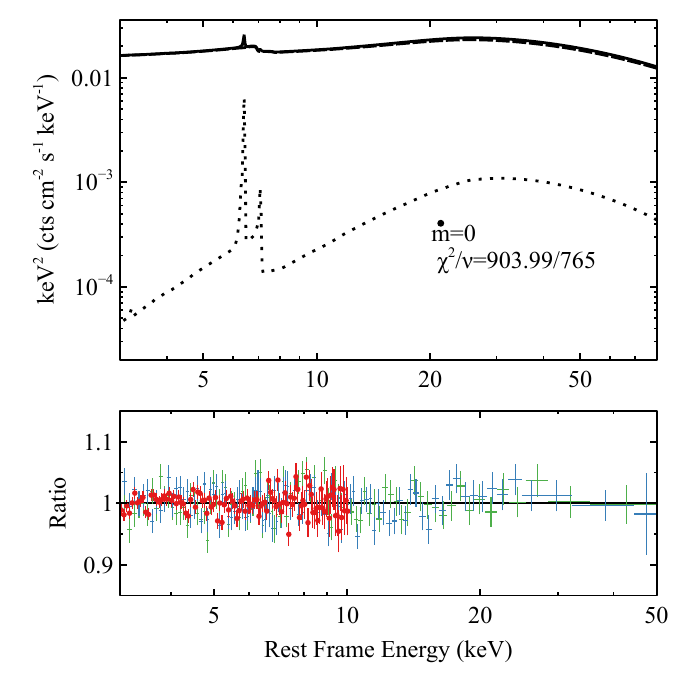}
    \caption{Top: best-fit razor-thin disc model ($\dot{m}=0$) for the pn and FPM spectra of \src. Solid: total model; dashed line: relativistic reflection component and coronal emission; dotted line: distant reflector. A zoom-in of the spectra in the iron emission band is in Fig.\,\ref{pic_fe}. Bottom: corresponding data/model ratio plot (red circles: pn; blue: FPMA; green: FPMB). The reflection fraction parameter is allowed to vary during the fit.}
    \label{pic_mdot0_fit}
\end{figure}

\begin{figure}
    \centering
    \includegraphics[width=\columnwidth]{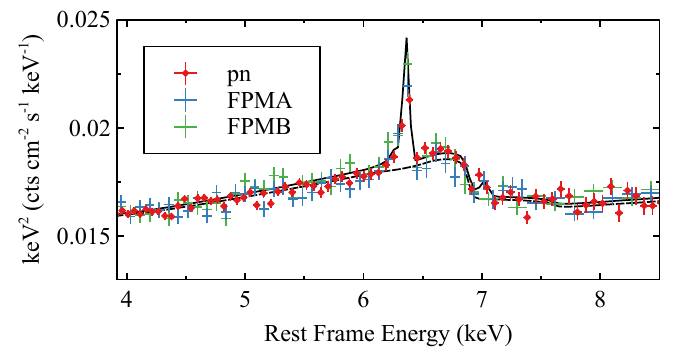}
    \caption{Best-fit razor-thin disc ($\dot{m}=0$) model overlaid with the data of \src\ (red circles: pn; blue: FPMA; green: FPMB). The dashed line shows the \texttt{relxilllpth\_nk} component.}
    \label{pic_fe}
\end{figure}

The best-fit razor-thin disc model parameters are shown in Table \,\ref{tab_ref_free}. The best-fit model is shown in Fig.\,\ref{pic_mdot0_fit}. The razor-thin disc model provides a very good fit to the data of \src\ with $\chi^{2}/\nu=903.99/765$. No significant residuals are found (see the lower panel of Fig.\,\ref{pic_mdot0_fit}). As shown in Fig.\,\ref{pic_fe}, the iron emission of \src\ has two component as suggested by previous analyses. The narrow emission line at 6.4\,keV is from the neutral reflector. The broad emission line is from the inner accretion disc.

\begin{figure}
    \centering
    \includegraphics[width=\columnwidth]{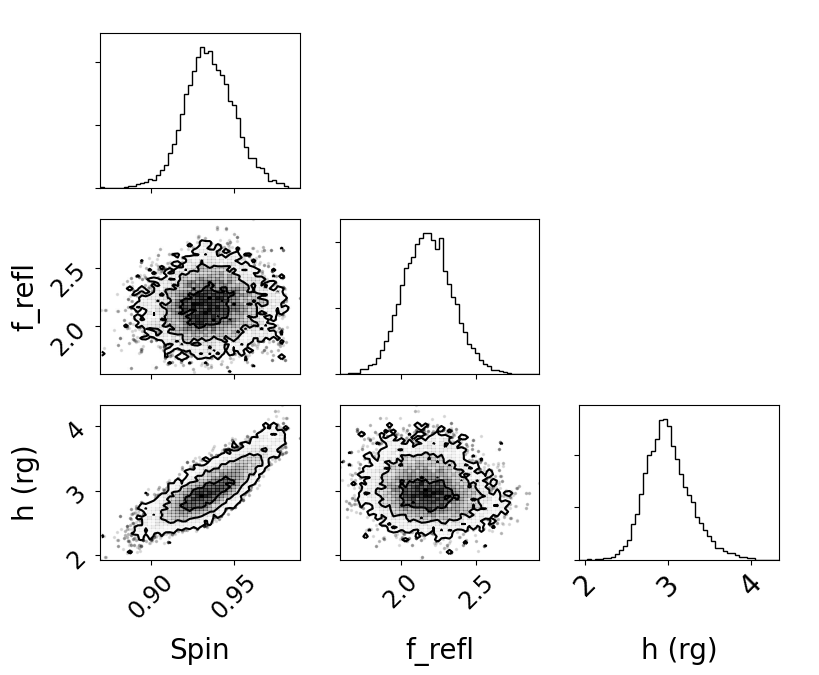}
    \caption{Output distributions for the MCMC analysis of the best-fit razor-thin disc model for \src.}
    \label{pic_mc0}
\end{figure}

We checked the constraints of all the parameters in this analysis by using the Markov chain Monte Carlo (MCMC) algorithm. We use 500 walkers with a \redch{total} length of 5000000, burning the first 20000. A convergence test has been conducted and the Gelman-Rubin scale-reduction factor $R<1.3$ for every parameter. No obvious degeneracy was found. In particular, we show the measurement uncertainties of $a_{*}$, $f_{\rm refl}$ and $h$ in Fig.\,\ref{pic_mc0}.

The best-fit razor-thin disc model infers a high BH spin of $0.94^{+0.02}_{-0.04}$ in \src, which is consistent with previous measurements by the same observations \citep[e.g. $a_{*}=0.91^{+0.06}_{-0.07}$, ][]{marinucci14}. Similar high BH spins were also achieved by other observations \citep[e.g.][]{young05}. Our model also suggests a compact coronal region above the accretion disc ($h\approx3r_{\rm g}$). The disc requires an iron abundance approximately 3.7 times the solar value similar to previous measurements \citep{marinucci14}. The total unabsorbed flux of the coronal emission and the relativistic disc reflection component is around $9.2\times10^{-11}$\,erg~cm$^{-2}$~s$^{-1}$ in the 3-50\,keV band. The distant, neutral reflection component has an unabsorbed flux of around $5\times10^{-13}$\,erg~cm$^{-2}$~s$^{-1}$ in the same energy band, which is only 0.6\% of the total X-ray flux.

So far, we have obtained a good fit using a razor-thin disc model. Our disc reflection parameters are consistent with previous results, although we do not include the soft X-ray data as in \citet{marinucci14, tripathi21}. We, therefore, continue focusing on only the Fe~K$\alpha$ emission and the Compton hump of \src\ in the following analysis. 

\begin{figure}
    \centering
    \includegraphics[width=\columnwidth]{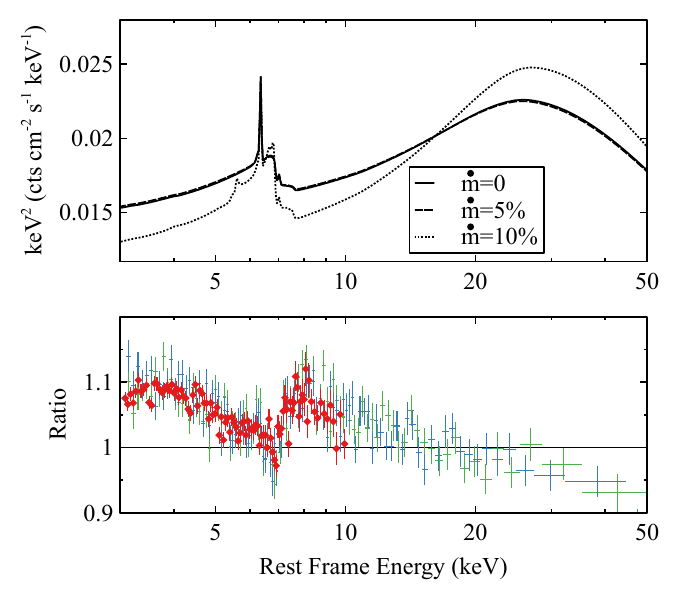}
    \caption{Top: the solid line shows the best-fit model based on the razor-thin disc model assuming $\dot{m}=0$. The dashed and dotted line shows the disc models for $\dot{m}=5\%,10\%$ respectively. \redch{Other parameters of the $\dot{m}=5\%,10\%$ models shown in this figure are set to be the same as the ones in the best-fit $\dot{m}=0$ model.} The difference between $\dot{m}=0$ and $\dot{m}=5\%$ models is small in the 3--50\,keV band. The $\dot{m}=10\%$ model, however, shows a significant different spectral shape--a double-peak Fe~K$\alpha$ emission is shown due to a flatter disc emissivity profile (see Fig.\,\ref{pic_q}). Bottom: data/model ratio plot for the spectra of \src\ fit to the $\dot{m}=10\%$ model in the upper panel.}
    \label{pic_2_o_0}
\end{figure}

\begin{table*}
    \centering
    \begin{tabular}{cccccccc}
    \hline\hline
      Models & Parameters & Units & $\dot{m}=0$ & $\dot{m}=5\%$ & $\dot{m}=10\%$ & $\dot{m}=20\%$ & $\dot{m}=30\%$ \\
      \hline
      \texttt{relxilllpth\_nk} & h & $r_{\rm g}$ & $3.0^{+0.7}_{-0.5}$ & $3.6^{+0.6}_{-0.8}$ & $2.4^{+0.3}_{-0.2}$ & $2.3^{+0.7}_{-0.3}$ & $2.4\pm0.4$\\
       & $i$ & deg & $39.8\pm1.7$ & $39.1^{+1.3}_{-1.0}$ & $40.4^{+0.6}_{-0.8}$ & $41.2\pm0.6$ & $37.9\pm1.7$ \\
       & $a_*$ & - & $0.94^{+0.02}_{-0.04}$ & $0.92^{+0.04}_{-0.05}$ & $0.95^{+0.02}_{-0.03}$ & $0.97\pm0.02$ & $0.97\pm0.02$ \\
       & $\log(\xi)$ & erg cm s$^{-1}$ & $1.69^{+0.06}_{-0.20}$ & $1.69^{+0.04}_{-0.07}$ & $1.70^{+0.04}_{-0.20}$ & $1.70^{+0.10}_{-0.20}$ & $1.69^{+0.06}_{-0.20}$ \\
       & $Z_{\rm Fe}$ & $Z_{\odot}$ & $3.7^{+0.5}_{-0.2}$ & $3.4^{+0.6}_{-0.3}$ & $3.5\pm0.2$ & $3.97^{+0.10}_{-0.08}$ & $3.7^{+0.2}_{-0.3}$ \\
       & $\Gamma$ & - & $1.94\pm0.02$ & $1.94\pm0.02$ & $1.95^{+0.02}_{-0.04}$ & $1.93\pm0.02$ & $1.94\pm0.02$ \\
       & Ecut & keV & $100\pm12$ & $106^{+12}_{-9}$ & $103\pm11$ & $97^{+8}_{-4}$ & $100^{+13}_{-10}$ \\
       & $f_{\rm refl}$ & - & $2.1^{+0.5}_{-0.4}$ & $1.25^{+0.10}_{-0.18}$ & $1.09^{+0.10}_{-0.20}$ & $0.84^{+0.17}_{-0.04}$ & $0.45^{+0.11}_{-0.07}$ \\
       & $\log(F_{\rm disc})$ & erg cm$^{-2}$ s$^{-1}$ & $-10.035\pm0.007$ & $-10.003^{+0.007}_{-0.004}$ & $-10.032^{+0.007}_{-0.009}$ & $-10.039\pm0.005$ & $-10.036\pm0.007$ \\
       \hline
       \texttt{xillver} & $\log(F_{\rm dis})$ & erg cm$^{-2}$ s$^{-1}$ & $-12.26\pm0.06$ & $-12.28\pm0.08$ & $-12.31^{+0.07}_{-0.04}$ & $-12.28^{+0.05}_{-0.03}$ & $-12.4\pm0.2$ \\
       \hline
       & $\chi^{2}/\nu$ & - & 903.99/765 & 904.44/765 & 899.41/765 & 900.58/765 & 907.73/765 \\
      \hline\hline
    \end{tabular}
    \caption{Best-fit parameters assuming different $\dot{m}$. $F_{\rm disc}$ and $F_{\rm dis}$ are the 3-50\,keV band flux of the disc reflection component and the distant reflector respectively.}
    \label{tab_ref_free}
\end{table*}

\subsection{Thin Disc Models with Finite Thickness} \label{chad}

To demonstrate how the thickness of the disc affects the shape of Fe~K$\alpha$ emission line in the reflection model, we show spectral models for different $\dot{m}$ in Fig.\,\ref{pic_2_o_0}. The black solid line shows the best-fit razor-thin disc model for \src. Then we only change the mass accretion rate parameter and keep other parameters the same. For instance, the dashed line shows the model for $\dot{m}=5\%$, which is almost consistent with the razor-thin disc model. Their difference is less than 5\% in the 3-50\,keV band. However, the broad Fe~K$\alpha$ emission line in the model starts showing a double-peak shape when $\dot{m}>10\%$. The $\dot{m}=10\%$ model shows a weaker red wing  than the $\dot{m}=0,5\%$ models. Because the region of the accretion disc around 5\,$r_{\rm g}$, which is less affected by gravitational redshift, is more illuminated when the disc is thicker. This leads to a flatter disc emissivity profile for $r<5$\,$r_{\rm g}$ as shown in Fig.\ref{pic_q} and thus changes the shape of the disc emission lines.  

We then apply the thin disc model with finite thickness to the same spectra analysed in Section\,\ref{m0}. The $\dot{m}$ parameter is set at 5\%, 10\%, 20\% and 30\% respectively\footnote{Note that the $\dot{m}$ parameter in \texttt{relxilllpth\_nk} is defined as $\dot{M}/\dot{M_{\rm Edd}}$, where $\dot{M}$ is the mass accretion rate and $L_{\rm Bol}=\epsilon\dot{M}c^{2}$. $\epsilon$ is the radiative efficiency, the value of which for \src\ is uncertain due to observational limit and Galactic obscuration \citep{raimundo12}. We do not further discuss the accretion efficiency of \src\ as this is beyond the purpose of this work.}. This model has the same free parameters as the razor-thin disc model introduced in Section\,\ref{m0}.

\begin{figure*}
    \centering
    \includegraphics[width=17cm]{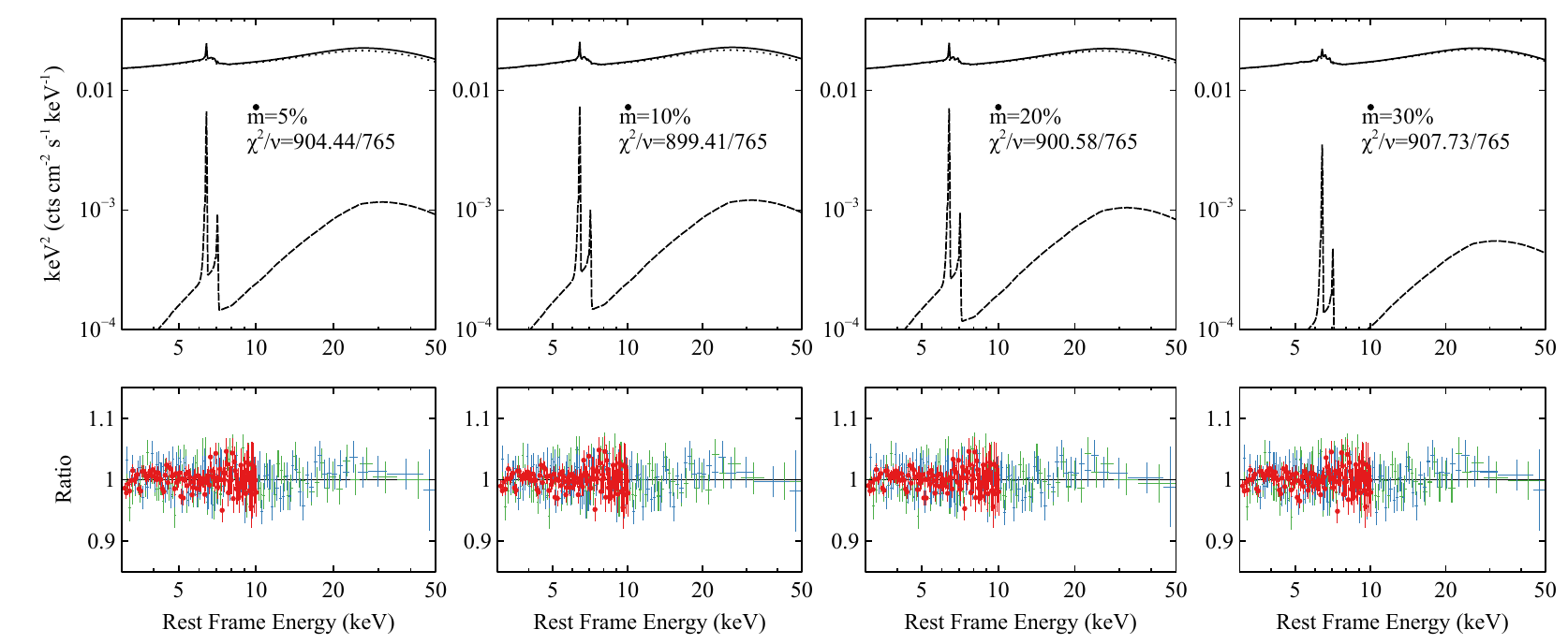}
    \caption{The same as Fig.\,\ref{pic_mdot0_fit} but for different $\dot{m}$. The reflection fraction parameter is allowed to vary during the fit. The $\dot{m}=30\%$ model provides a slightly worse fit to the data than other models ($\Delta\chi^{2}\approx3-8$ and the same number of free parameters).}
    \label{pic_thin_disc}
\end{figure*}

Best-fit models are shown in Fig.\,\ref{pic_thin_disc} and best-fit parameters are shown in Table\,\ref{tab_ref_free}. These thin disc models with finite thickness provide similar good fits to the data of \src\ as the razor-thin disc model. The model with the highest mass accretion rate ($\dot{m}=30\%$) offers a slightly worse fit than others by $\Delta\chi^{2}=3-8$ with the same number of free parameters. Although disc thickness is considered in these models, most of the parameters have consistent best-fit values as the razor-thin disc model, e.g. the photon index and the high-energy cutoff of the continuum emission, the iron abundance and the ionisation of the disc. We conducted similar MCMC analysis as in Section \ref{m0}. The output distributions of the same key parameters in Fig.\,\ref{pic_mc0} are shown in Fig.\,A\ref{pic_mn0}.

We note that a slightly lower $h$ is needed to fit the data when a higher $\dot{m}$ is assumed, although the values are consistent within their 90\% confidence ranges. This is because the emissivity profile is flatter when and the disc becomes thicker at high $\dot{m}$ (see Fig\,\ref{pic_q} and Fig.\,\ref{pic_2_o_0}). A more compact coronal region is required in the model. So the inner disc produces more reflected light to fit the observed red wing of the broad Fe~K$\alpha$ emission line. Similarly, a slightly higher BH spin is suggested by the high-$\dot{m}$ models, although they are all consistent within their 90\% confidence ranges. A higher BH spin at high $\dot{m}$ indicates a smaller inner disc radius, which also produces an extended red wing in the broad Fe~K$\alpha$ emission to fit the data.

\begin{figure}
    \centering
    \includegraphics[width=\columnwidth]{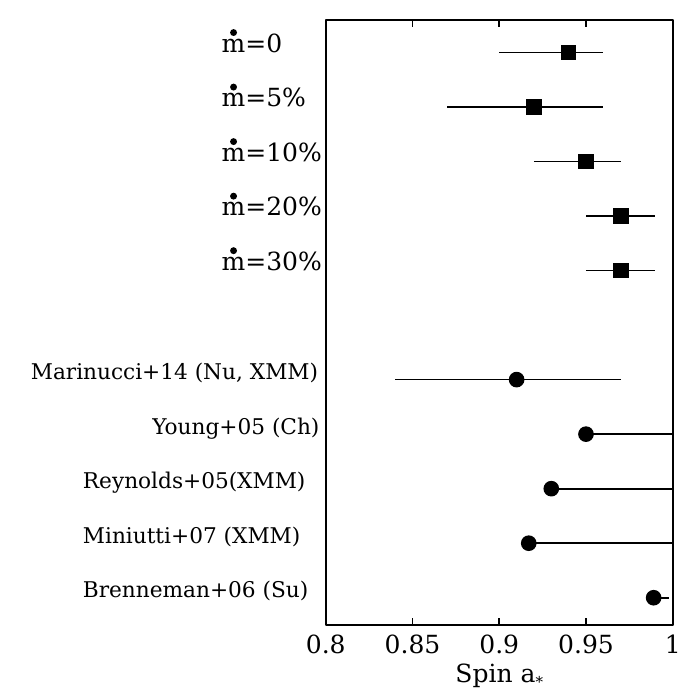}
    \caption{Best-fit spin parameters in this work (squares) in comparison with previous measurements (circles) in the literature \citep{young05,reynolds05,brenneman06,miniutti07,marinucci14}. The data analysed in their work are noted behind the references. Nu: \nustar; XMM: \xmm; Ch: \chandra\ grating data; Su: \suzaku.}
    \label{pic_spin}
\end{figure}

In particular, we compare the spin measurements obtained in this work (black squares) with previous results (black circles) based on razor-thin disc models in Fig.\,\ref{pic_spin}. Among previous work, \citet{young05} analysed the grating data of \src\ from \chandra\ while others all considered CCD-resolution data. The highest value of BH spin was achieved by \citet{brenneman06} where \suzaku\ data of \src\ were analysed and a close-to-maximum BH spin of $a_{\*}=0.989^{+0.009}_{-0.002}$ was obtained. It is important to note that \citet{brenneman06} considered a relativistic line model instead of a full disc reflection spectral model. A lower BH spin of $a_{*}=0.91^{+0.06}_{-0.07}$ was achieved by \citet{marinucci14}, where the combination of \texttt{relconv * xillver} was used instead of \texttt{relxill}. \citet{tripathi20} studied the difference between the \texttt{relconv * xillver} and \texttt{relxill} models for \src. The latter model in particular considers the angle dependence of the emission at different regions of the disc while the former does not. This effect may play an important role in reflection modelling and is considered in our thin disc model. Nevertheless, our spin measurements at all $\dot{m}$ are consistent with previous results in \citet{young05, reynolds05, miniutti07, marinucci14}.

In summary, the lower limit of the BH spin in \src\ is 0.87 for $\dot{m}=5\%$ and the higher limit is 0.99 for $\dot{m}=20\%$. The mass accretion rate of \src\ is uncertain. We, therefore, conclude that the 90\% confidence range of the BH spin in \src\ is 0.87--0.99 after taking the thickness of the disc into consideration. This result is consistent with previous measurements based on razor-thin disc models. Besides, a slightly higher $a_{*}$ and a more compact coronal region is needed in a higher-$\dot{m}$ disc model to fit the red wing of the observed broad Fe~K$\alpha$ emission. We will present spectral simulations for further discussion in the following paper of this series.

\section{Reflection Fraction in the thin disc model} \label{link}

We have obtained a good fit for the spectra of \src\ using a thin disc model with finite thickness. A grid of $\dot{m}$ up to 30\% is considered. They provide similar good fits to the data and achieve consistent BH spin measurements as before. 

\begin{figure}
    \centering
    \includegraphics[width=\columnwidth]{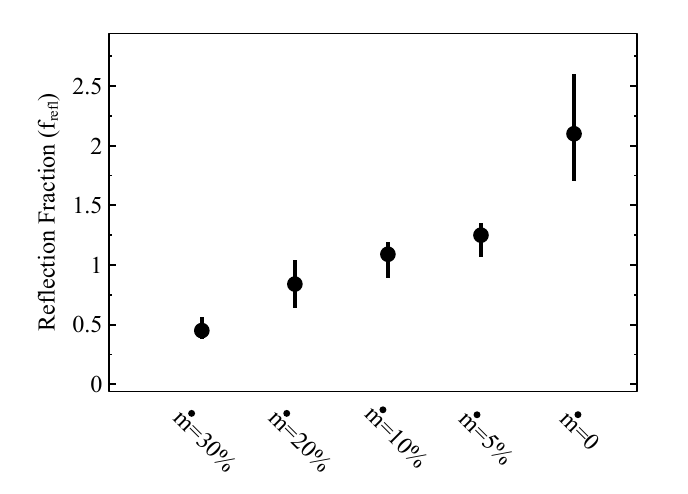}
    \caption{Best-fit reflection fraction parameters when different $\dot{m}$ is used in the model. }
    \label{pic_frefl_obs}
\end{figure}

A free reflection fraction parameter is used in the models. \redch{The reflection fraction parameter in \texttt{relxilllpth\_nk} is defined as the ratio of the coronal intensity illuminating the disk to the coronal intensity that reaches the observer\footnote{This parameter is referred to as the system reflection fraction in \citet{ingram19}.} \citep{dauser16}. The value of this parameter can easily reach a very high value significantly larger than 1 due to the light bending effects in a compact coronal region.} The best-fit reflection fraction parameter shows an anti-correlation with the assumed $\dot{m}$ values (see Fig.\,\ref{pic_frefl_obs} or Table \ref{tab_ref_free}): the razor-thin disc model has $f_{\rm refl}\approx2.1$ and the $\dot{m}=30\%$ model has $f_{\rm refl}\approx0.45$. This discrepancy needs further discussion and calculations as following.

We first fit the data of \src\ to thin disc reflection models where $f_{\rm refl}$ is calculated consistently according to the height of the lamppost corona and the thickness of the disc. We then compare the fits of such models to our previous fits with a free $f_{\rm refl}$ parameter. In the end, we calculate the theoretical values of $f_{\rm refl}$ in the thin disc model and discuss our results. 

\subsection{Spectral Models with Consistent Reflection Fraction}

To adapt consistent values of $f_{\rm refl}$, we apply the same model to the data of \src\ in XSPEC as in Section\,\ref{m0}. The only difference is that \texttt{relxilllpth\_nk} calculates $f_{\rm refl}$ according to the height of the lamppost corona and the thickness of the disc in the following analysis. This is achieved by numerical ray-tracing techniques. 

\begin{figure}
    \centering
    \includegraphics[width=\columnwidth]{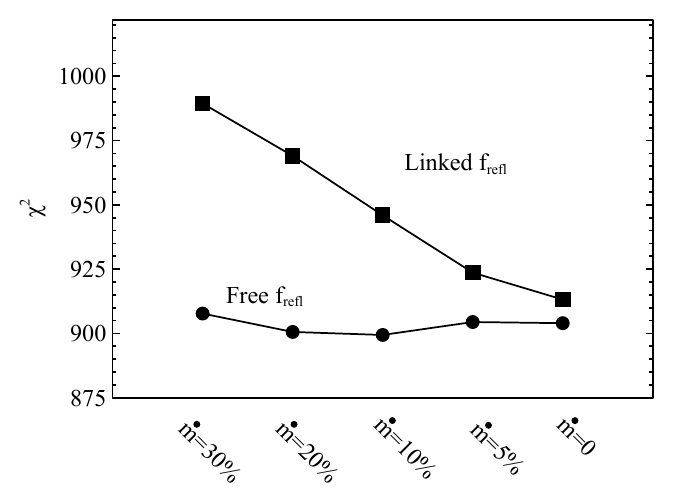}
    \caption{$\chi^{2}$ vs. $\dot{m}$. The circles show the $\chi^{2}$ values of models with a free $f_{\rm refl}$ parameter. The squares show the values of models where $f_{\rm refl}$ is calculated consistently for a thin disc illuminated by the lamppost corona.}
    \label{pic_chi2}
\end{figure}

\begin{figure*}
    \centering
    \includegraphics{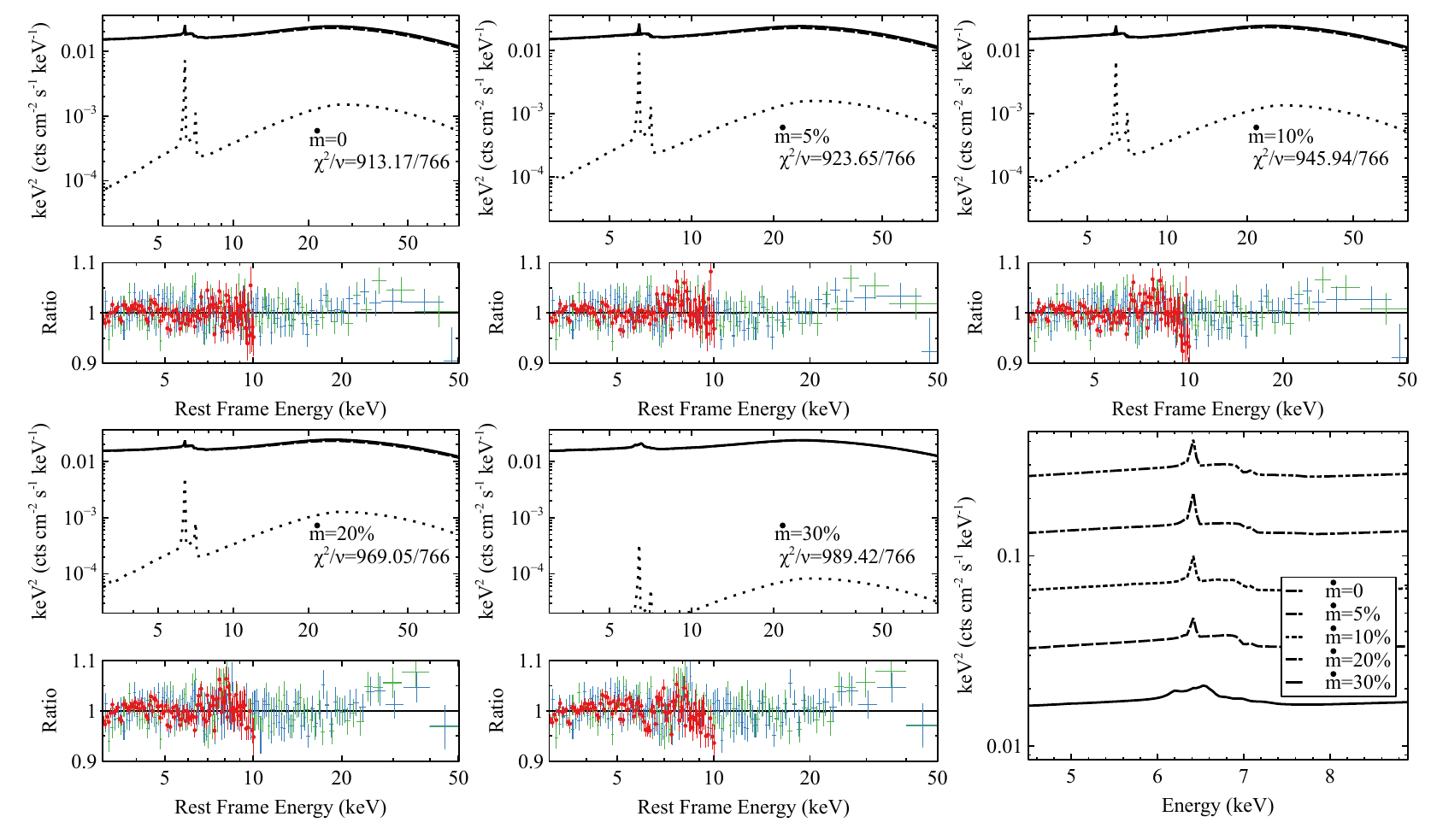}
    \caption{Same as Fig.\,\ref{pic_mdot0_fit} but with a linked $f_{\rm refl}$ parameter. Significant residuals are seen in the iron emission band and the hard X-ray band, e.g. >20\,keV when $\dot{m}>10\%$. The bottom right panel shows the zoom-in of the best-fit models in the iron emission band. The models are shifted vertically for clarification. See text for more details.}
    \label{pic_fix}
\end{figure*}

Best-fit values of thin disc models with linked $f_{\rm refl}$ are shown in Table\,\ref{tab_ref_link} and corresponding models are shown in Fig.\,\ref{pic_fix}. The $\chi^{2}$ distribution against $\dot{m}$ is shown in Fig.\,\ref{pic_chi2}. 

After including consistent values of $f_{\rm refl}$, we find that the models provide worse fits to the data than the ones in Section\,\ref{free}. For example, the razor-thin disc model with linked $f_{\rm refl}$ provides a worse fit than the one with a free $f_{\rm refl}$ parameter by $\Delta\chi^{2}=10$. The former has one fewer free parameter, $f_{\rm refl}$, than the latter. As shown in Fig.\,\ref{pic_fix}, positive residuals above 20\,keV are seen. The razor-thin disc model with linked $f_{\rm refl}$ cannot fit the Compton hump of the disc reflection component so well as the one with free $f_{\rm refl}$, although the goodness of the fit is acceptable. 

As shown in Fig.\,\ref{pic_chi2}, $\chi^{2}$ is particularly high at large $\dot{m}$ when a linked $f_{\rm refl}$ parameter is used. When $\dot{m}=30\%$, the model fails to fit not only the hard X-ray band but also the iron emission band. A low BH spin of $a_{*}=0.60$ is suggested by this fit. It is important to note that this model provides a much worse fit to the data than other models by $\Delta\chi^{2}\approx90$. We, therefore, reject this measurement. The inferred low BH spin in this fit suggests that a large inner disc radius is needed in the model to produce the right amount of disc reflection in the model. However, by doing so, the model fails to reproduce the observed Fe~K$\alpha$ line shape as shown in the right bottom panel of Fig.\,\ref{pic_fix}.

\begin{table*}
    \centering
    \begin{tabular}{cccccccc}
    \hline\hline
      Models & Parameters & Units & $\dot{m}=0$ & $\dot{m}=5\%$ & $\dot{m}=10\%$ & $\dot{m}=20\%$ & $\dot{m}=30\%$ \\
      \hline
      \texttt{relxilllpth\_nk} & h & $r_{\rm g}$ & $3.2^{+0.3}_{-0.4}$ & $2.6\pm0.3$ & $2.0^{+0.3}_{-0.2}$ & $2.3^{+0.3}_{-0.2}$ & $2.0\pm0.2$  \\
       & $i$ & deg & $38.8^{+0.7}_{-1.2}$ & $41.0^{+0.4}_{-1.6}$ & $40.7^{+0.8}_{-0.9}$ & $39.2^{+0.7}_{-0.2}$ & $42^{+3}_{-2}$ \\  
       & $a_*$ & - & $0.92^{+0.05}_{-0.04}$ & $0.95^{+0.04}_{-0.02}$  & $0.983^{+0.008}_{-0.010}$ & $0.92\pm0.02$ & $0.60^{+0.08}_{-0.02}$  \\
       & $\log(\xi)$ & erg cm s$^{-1}$ & $1.69^{+0.07}_{-0.13}$ & $1.69^{+0.10}_{-0.03}$ & $1.69^{+0.02}_{-0.03}$ & $1.69^{+0.02}_{-0.03}$ & $1.69\pm0.03$ \\
       & $Z_{\rm Fe}$ & $Z_{\odot}$ & $2.5\pm0.2$ & $2.8\pm0.2$  & $2.6^{+0.4}_{-0.3}$ & $2.15^{+0.12}_{-0.14}$ & $2.1\pm0.3$\\
       & $\Gamma$ & - & $2.030^{+0.010}_{-0.008}$ & $2.025^{+0.014}_{-0.011}$  & $2.080^{+0.010}_{-0.019}$ & $2.054^{+0.012}_{-0.002}$ & $2.049\pm0.002$ \\
       & Ecut & keV  & $123^{+7}_{-11}$ & $111^{+3}_{-5}$ & $125^{+12}_{-10}$ & $135^{+5}_{-13}$ & $124\pm16$\\
       & $f_{\rm refl}$ & - & $l$ & $l$  & $l$ & $l$ & $l$ \\
       & $\log(F_{\rm disc})$ & erg cm$^{-2}$ s$^{-1}$ & $-10.022\pm0.007$ & $-10.029^{+0.007}_{-0.010}$  & $-10.0243^{+0.007}_{-0.006}$  & $-10.024\pm0.008$  & $-10.012\pm0.005$ \\
       \hline
       \texttt{xillver} & $\log(F_{\rm dis})$ & erg cm$^{-2}$ s$^{-1}$ & $-12.22\pm0.04$ & $-12.17\pm0.04$  & $-12.23^{+0.05}_{-0.04}$  & $-12.31^{+0.05}_{-0.04}$ & <-13.2 \\
       \hline
       & $\chi^{2}/\nu$ & - & 913.17/766 & 923.65/766 & 945.94/766 & 969.05/766 & 989.42/766\\
      \hline\hline
    \end{tabular}
    \caption{Best-fit parameters for different $\dot{m}$. $f_{\rm refl}$ is calculated consistently by given $a_*$ and $h$ in the model. $F_{\rm disc}$ and $F_{\rm dis}$ are the 3-50\,keV band flux of the disc reflection component and the distant reflector respectively. See text for more details.}
    \label{tab_ref_link}
\end{table*}

So far, we found that the thin disc model provides significantly worse fits to the data of \src\ when considering lamppost values of $f_{\rm refl}$. Among the fits with linked $f_{\rm refl}$ parameters, the razor-thin disc model provides the best fit to the data, although its fit is still worse than the one with a free $f_{\rm refl}$ parameter. 

In the following section, we argue that the reason why models with a linked $f_{\rm refl}$ parameter provide worse fits to the data is that they overestimate the reflection fraction in \src. 


\subsection{Reflection Fraction of a Thin Disc Illuminated by the Lamppost Corona} \label{frefl}

\begin{figure*}
    \centering
    \includegraphics[width=\textwidth]{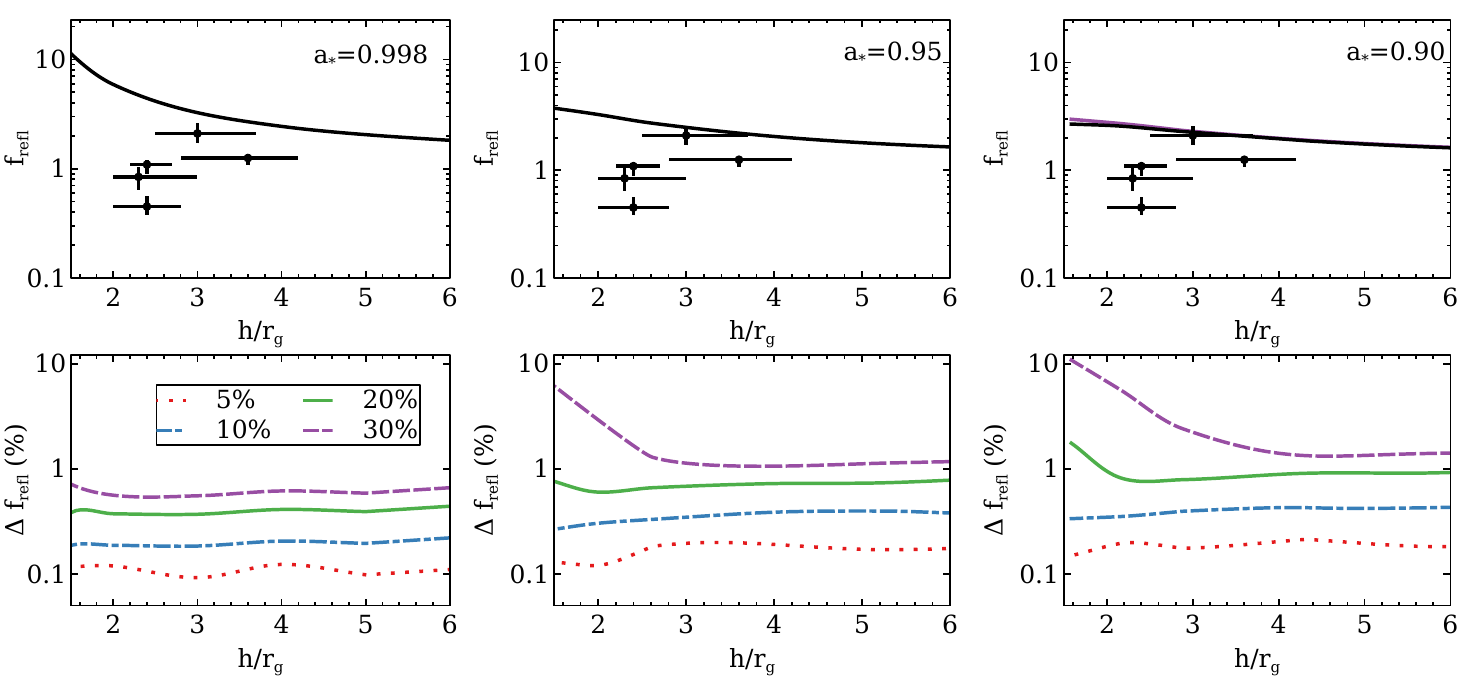}
    \caption{Top panels: the black solid lines show $f_{\rm refl}$ of a razor-thin disc ($\dot{m}=0$) illuminated by the lamppost corona. The three panels are for $a_{*}=0.998, 0.95, 0.90$ from left to right. The crosses show the observed $h$ and $f_{\rm refl}$ for $\dot{m}=0,5\%,10\%,20\%,30\%$ from top to bottom respectively. Bottom panels: difference in $f_{\rm refl}$ for different $\dot{m}$. For instance, the increase of $f_{\rm refl}$ is expected to be less than 1\% when the thickness of the disc is considered for maximum BH spin. Significant difference up to a few per cent is only seen when the mass accretion rate is high, spin has an intermediate value and the corona is very compact, e.g. $a_{*}=0.90$, $\dot{m}=30\%$ and $h<3r_{\rm g}$. We show the reflection fraction of a disc with $\dot{m}=30\%$ in purple in the top right panel in comparison. Note that the best-fit values of $f_{\rm refl}$ are all lower than the expected reflection fraction by given $h$, which explains why models with a consistent $f_{\rm refl}$ parameter provide worse fits to the data of \src\ than the ones with a free $f_{\rm refl}$ parameter. Modifications to the simple lamppost corona might be needed when a thin disc model with finite thickness is considered.}
    \label{pic_frefl_lp}
\end{figure*}

The reflection fraction parameter in the lamppost geometry was calculated in detail by \citet{dauser16} where a razor-thin disc was used. We show three examples for different BH spins in Fig.\,\ref{pic_frefl_lp}. The reflection fraction of the disc increases with decreasing $h$. Because the inner disc is more illuminated when the corona is closer. The spin of the BH also plays an important role: the disc has a lower reflection fraction for an intermediate spin than a near maximum spin. The theoretical values of the reflection fraction parameter in the lamppost model can easily go over 1 due to the strong light-bending effects near a BH. Only when $h$ is very large, the flux that reaches the disc and infinity becomes similar and the reflection fraction approaches 1. 

We show the difference of reflection fraction in disc models with infinite small height and finite thickness in the lower panels of Fig.\,\ref{pic_frefl_lp}. When the disc becomes thicker at higher $\dot{m}$, a higher reflection fraction is expected. However, the overall increase of reflection fraction is less than 10\%. When the central BH has a maximum spin, the difference is as small as less than 1\%. 

Fig.\,\ref{pic_frefl_lp} also shows that the reflection fraction of the disc is less affected by  mass accretion rate when the BH spin is high. Because $\dot{m}$ has a smaller impact on the thickness of the disc when the BH spin is high (see Fig.\,\ref{pic_shape}). Only when 1) the BH has a modest spin; 2) the mass accretion rate of the disc is very high; 3) the corona is very compact, the difference between the reflection fraction in two models is larger than 1\%. For example, we show the reflection fraction for $a_{*}=0.90$ and $\dot{m}=30\%$ in purple in the top right panel of Fig.\,\ref{pic_frefl_lp}. The difference between the black and purple curves is up to 10\% only when the corona is very close to the event horizon of the BH.

We show the observed values of $h$ and $f_{\rm refl}$ obtained in Section \ref{free} in Fig.\,\ref{pic_frefl_lp}. Note that the best-fit BH spin of \src\ are 0.92--0.97 (see Table\,\ref{tab_ref_free}). The observed values of  $f_{\rm refl}$ are all lower than the theoretical values in the lamppost geometry. Only when $a_{*}=0.90$, the expected reflection fraction by the razor thick disc model is consistent with the observed value. The difference between observed and theoretical values of $f_{\rm refl}$ is larger when the thin disc model with thickness is considered. The overestimation of reflection fraction in the lamppost model explains the residuals in the hard X-ray band where Compton hump is (see Fig.\,\ref{pic_fix}). 

In summary, we apply a thin disc model, which takes the thickness of the disc into consideration and calculates consistent $f_{\rm refl}$ assuming the lamppost geometry, to the data of \src. We find that the razor-thin disc model provides a better fit to the data than other models with finite thickness. Because the lamppost geometry overestimates the reflection fraction of the disc when the thickness of the disc is considered. Although this razor-thin disc model with linked $f_{\rm refl}$ offers a slightly worse fit than the models with free $f_{\rm refl}$ by $\Delta\chi^{2}=10$, their BH spin measurements are still consistent.

\section{Other Systematic Uncertainties in the Disc Reflection Model}

\red{In this work, we study the effects of thin disc geometry on disc reflection spectral modelling. Our analysis is based on the \xmm\ and \nustar\ observing campaign of the well-studied X-ray bright AGN \src. We ignore the soft X-ray data below 3\,keV to focus only on the broad Fe K emission and the Compton hump in the spectra, which are the prominent features of disc reflected emission.} \red{A certain geometry has to be chosen for the corona in our model as it determines the emissivity profile of the disc as well as the thickness of the disc \citep[e.g.][]{wilkins12,gonzalez17}. One has to consider a specific coronal geometry to separately study the systematic effects of disc thickness in the reflection model. We, therefore, consider the best-understood lamppost geometry for the coronal region in our model to start with. }

\red{By fitting the spectra of \src, we find that the thin disc model provides a similarly good fit to the data and a similar constraint on BH spin as the razor-thin disc model. However, as argued in \citet{dauser16} and Section\,\ref{frefl}, the reflection fraction of the disc is also an indicator of the geometry of the innermost accretion region. This parameter is expected to increase in the thin disc model compared to the simplified razor-thin disc model. But the increase of reflection fraction is usually less 10\% unless in extreme cases (see Fig.\,\ref{pic_frefl_lp}), while the observed values of this parameter in the thin disc model are significantly lower than the expectations of the lamppost model.} 

\red{The discrepancy in the observed and expected values of disc reflection fraction leads to the discussion in this section: is the lamppost geometry valid? In particular, is the lamppost geometry applicable to the X-ray data of \src? Are there other systematic uncertainties in the disc reflection model, e.g. the properties of the disc in \src, which we did not take into account? We also refer interested readers to the latest review by \citet{bambi21} for a similar but more detailed discussion.}

\subsection{Is the Lamppost Geometry Valid in MCG-06-30-15?} \label{olet}


\red{Following the same approach in \citet{taylor18a}, we implement the best-understood lamppost geometry \citep{martocchia96} in our thin disc model for simplicity in Section \ref{free}.}
\red{The best-fit thin disc mode suggests a compact coronal region of 2--4\,$r_{\rm g}$ and a relatively lower Ecut\footnote{The temperature of the thermal Comptonisation corona is usually half or a third of the value of Ecut. \second{Ecut=100\,keV corresponds to a temperature of $kT_{\rm e}=30-50$\,keV.}} at around 100\,keV (see Table\,\ref{tab_ref_free}) in comparison with the sample in \citet{fabian15}. If the coronal region is stationary and very compact in \src\ as suggested by the lamppost model, one might need to consider the strong relativistic effects, e.g. gravitational redshifts, on the observed values of coronal temperatures and luminosities \citep{niedzwiecki16}.} \second{The relation between the observed luminosity and the source-frame luminosity is $L_{\rm obs} = g^4 \ell L_{\rm s}$ (ignoring the small cosmological redshift), where $g$ is the redshift factor and $\ell$ is the lensing factor. For the lamppost model, $g = \sqrt{1 - 2 h r_{\rm g}/(h^2 + a^2_* r^2_{\rm g})}$ \citep{bambi21} and $\ell$ can be found, for instance, in Fig.~3 in \citet{ingram19}. For $h = 4 r_{\rm g}$ and an observed X-ray luminosity of 4\% of its Eddington limit for \src\ (see Section\,\ref{disk_ge} for detailed calculations of X-ray luminosities), we have $g \approx 0.73$, $\ell \approx 0.4$, and $L_{\rm s} \approx 0.35 L_{\rm Edd}$. The temperature of the corona in \src\ is around $kT_{\rm e}/$g$\approx$20--30/$g$=40--70\,keV. On the other hand, for $h = 2 r_{\rm g}$, $g \approx 0.45$, $\ell \approx 0.2$, $L_{\rm s} \approx 5 L_{\rm Edd}$ and we find a super-Eddington luminosity in the rest-frame of the source. Coronae very close to a black hole can also be affected by the runaway $e^\pm$ pair production \citep{fabian15,niedzwiecki16}.} 

\second{However, we note that our value of $h$ is to be taken with caution as inferred within the ideal lamppost coronal geometry (point-like source, exactly along the rotational axis of the black holes, with isotropic emission). A realistic corona is extended and non-isotropic and this motivates the development of more physically consistent models \citep[see ,e.g., the discussion in][]{niedzwiecki16,bambi21}.} \red{A more physical model might be required for \src\ especially when the thin disc model is considered. For example, we find a negative correlation between the inferred disc reflection fraction parameter and the value of $\dot{m}$ used in the model, which is driven by the spectral fitting of the broad Fe~K emission in the data. The inferred reflection fraction parameter of the razor-thin disc model is consistent with the expected value for $a_{*}=0.9$. But all values are lower than the expected values in the thin disc models, making the thin disc models with the reflection fraction parameter consistently calculated for the lamppost geometry fail to fit the data so well as the ones with a free reflection fraction parameter (see Fig.\,\ref{pic_chi2} for comparison of $\chi^{2}$).}

\red{Based on the discrepancy between the observed and expected values of reflection fraction in the thin disc model for \src, a more complex coronal region with a more extended shape, e.g. an outflow/jet-like corona \citep{wilkins14}, may exist in \src. An outflowing corona explains the overestimation of reflection fraction in the lamppost model. Unfortunately, complex coronal geometries have not been implemented in the thin disc model yet due to the complexity of calculations.}

\red{The X-ray time-series data analysis of \src\ supports a similar conclusion of a potentially more complex coronal geometry in \src. The reduced Root-Mean-Square (RMS) variability in the iron emission band of \src\ was noticed in its \textit{ASCA} data \citep{matsumoto03}. No significant evidence of correlation among the variability of the blue and red wings of the Fe~K emission line and the primary X-ray continuum was found either \citep{matsumoto03}.  Similar conclusions were found through the Principal Component Analysis of the \xmm\ observations of this object \citep{parker14x}. The disc reflection component is responsible for less than 1\% of the X-ray variability. Although the light-bending model \citep{miniutti03} explains the weak variability of the reflection component, the lack of highly variable reflection in \src\ leads to the reduced correlated variability in the iron emission band \citep{kara14}. Not all disc reflection is correlated with the variability of the corona in \src\ at all Fourier frequencies, including the frequency range where Fe~K reverberation lags are usually found in other AGN. Thus no obvious evidence of reverberation lag has been detected\footnote{\citet{chainakun22} tried to find a correlation between the amplitudes of Fe~K reverberation lags, BH masses, X-ray variability and some other parameters in a sample of 22 reverberating AGN based on neural networks. No assumptions for coronal geometries were made in this work. The neural network output suggested that the fractional excess variance of \src\ might be too low for a BH as small as the one in \src\ to show reverberation lags.} in \src\ \citep{kara14}.}

\red{One possible reason for the insignificant evidence of Fe~K lag and reduced covariance variability in the iron emission band might be a moving corona with changing line-of-sight velocity \citep{matsumoto03}. When the corona moves towards the observer with a speed of a significant fraction of the light speed, the primary continuum emission is highly beamed and boosted due to relativistic effects. Meanwhile, the disc is less illuminated than in the stationary lamppost geometry. A lower reflection fraction and disconnection between the Fe~K emission and the primary continuum are thus expected. Calculations for complex coronal geometries in combination with the thin disc model are needed for \src. They may help explain both the observed values of reflection fraction in its energy spectra and the decrease of correlated variability in its covariance spectra.}





\subsection{The Geometry of the Inner Accretion Disc of MCG-6-30-15} \label{disk_ge}

\begin{figure}
    \centering
    \includegraphics[width=8cm]{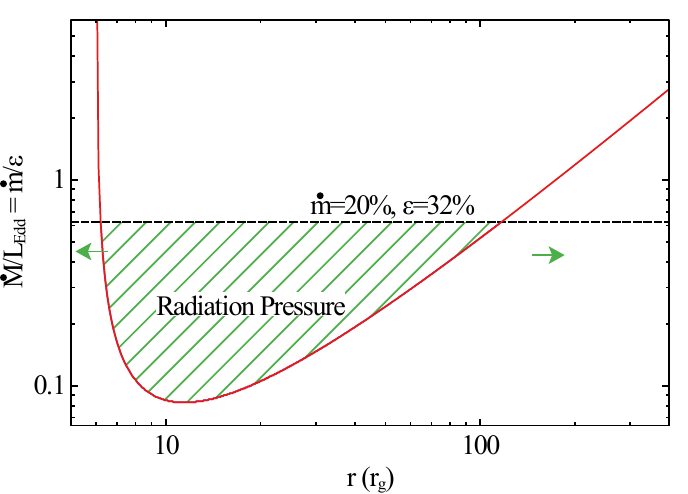}
    \caption{\red{The critical mass accretion ($\dot{m}/\epsilon$) where gas and radiation pressure equals as a function radius $r$ in the standard thin disc model \citep[][in red solid curve]{shakura73}. We use $M_{\rm BH}=1.6\times10^{6}$\,$M_{\odot}$ \citep{bentz16} and $\alpha=0.1$. Note that this solution is for the flat space time with the inner disc radius at $6$\,$r_{\rm g}$. This function has a lowest point at 11.44\,$r_{\rm g}$ corresponding to $\dot{m}/\epsilon=0.083$. When $\dot{m}/\epsilon$ is lower than this value, there is no radiation pressure-dominated region in the disc. When $\dot{m}/\epsilon$ is higher than 0.083, the radiation pressure-dominated region expands both outwards and inwards. Assuming the highest accretion efficiency of $\epsilon=32\%$ for the maximum BH spin and $\dot{m}=20\%$ for \src, the radiation pressure will dominate in the region of $r=6.1-117$\,$r_{\rm g}$ (the shaded region). A higher $\dot{m}$ or a lower $\epsilon$ would lead to an even larger radiation pressure-dominated region in \src.}}
    \label{pic_gas}
\end{figure}

\red{In the model presented in this work, we assume the thickness of the disc to be scaled with the pressure scale height in the radiation pressure-dominated region of the standard geometrically thin, optically thick disc \citep{shakura73}. In this disc theory, the radiation pressure-dominated region exists in the innermost accretion region of a disc when the mass accretion rate is higher than a critical value and is thermally unstable \citep[e.g.][]{lightman74,honma91}.} \red{Other approximations have been made in \citet{shakura73} too: the thin disc model was based on the flat spacetime but very similar to the relativistic solutions \citep{novikov73}; the effects of the energy transformation from the disc to the corona were not considered either in this model \citep{haardt91,haardt93,svensson94}.} 

\red{The thickness profiles $z/r$ for different BH spins used in this work are given in Equation \ref{Hva} and Fig.\,\ref{pic_shape}. 
BH spins change the $z/r$ profile by affecting $r_{\rm isco}$ and $\epsilon$. Our model also assumes the radiation pressure dominates the region where disc reflection mostly comes from. We review the radiation and gas pressure solutions in \citet{shakura73} and argued that the radiation pressure solutions are a good approximation. To do so, we calculate the the radius where gas pressure ($P_{\rm gas}$) and radiation pressure ($P_{\rm rad}$) are equal as a function of mass accretion rate $\dot{M}c^{2}/L_{\rm Edd}=\dot{m}/\epsilon$. The results are given in Fig.\,\ref{pic_gas}. The calculations are based on Equation 29 in \citet{svensson94}. We assume a viscosity parameter of $\alpha=0.1$, a BH mass of $1.6\times10^{6}M_{\odot}$ as in \src\ \citep{bentz16} and zero coronal power $f=0$ to recover the solutions in \citet{shakura73}. The inner radius of the disc is assumed to be 6\,$r_{\rm g}$ in \citet{shakura73}.}

\red{When $\dot{m}/\epsilon$ is below 0.083, there is no solution where $P_{\rm gas}=P_{\rm rad}$ in the disc as shown in Fig.\,\ref{pic_gas} and the whole disc is dominated by $P_{\rm gas}$. At $\dot{m}/\epsilon=0.083$, $P_{\rm rad}$-dominated region appears at 11.4\,$r_{\rm g}$. When $\dot{m}/\epsilon$ is higher 0.083, the $P_{\rm rad}$-dominated region expands both inwards and outwards.}

\red{We estimate the size of the $P_{\rm rad}$-dominated region in \src\ by starting with the estimation of its Eddington ratio. The unabosrbed flux of \src\ is $4.6\times10^{-11}$ ($3.5\times10^{-11}$) \ergs\ in the 2--10\,keV (3--10\,keV) band, corresponding a luminosity of $8.2\times10^{44}$\,erg\,s$^{-1}$ which is 4\% of the Eddington luminosity for $1.6\times10^{6}M_{\odot}$. Assuming a very low bolometric correction factor of 5 \citep{vasudevan07,netzer19}, we estimate the Eddington ratio of \src\ to be at least 20\%. For sources that are accreting at a significant fraction of the Eddington limit, a higher bolometric correction factor for the 2--10\,keV luminosity might be needed, which would only lead to a higher Eddington ratio estimation. $\dot{M}/\dot{M_{\rm Edd}}$ is approximately $L_{\rm Bol}/L_{\rm Edd}$ assuming that the accretion efficiency is the same at $\dot{M_{\rm Edd}}$ and a significant fraction of $\dot{M_{\rm Edd}}$. To estimate the lower limit of $\dot{m}/\epsilon$ for \src, we consider the highest value of $\epsilon$ at 32\% for the maximum BH spin. The horizontal dashed line in Fig.\,\ref{pic_gas} shows that the lower limit of the size of the $P_{\rm rad}$-dominated region in \src\ (6.1-117\,$r_{\rm g}$). In this calculation, we consider the lowest bolometric correction factor and the highest radiative efficiency. A higher bolometric correction factor or a lower radiative efficiency would lead to a larger $P_{\rm rad}$-dominated region size in \src. In conclusion, we find that the radiation-pressure solutions are a good approximation of the inner disc region in \src\ where most reflection originates in according to the standard thin accretion disc model. A more accurate model, e.g. for the Kerr spacetime or inclusion of the outer $P_{\rm gas}$-dominated region, introduces higher order corrections.}
 
\red{Recently, optically thick outflows have also been proposed to be the origin of the broad Fe~K emission in AGN. In AGN with a high Eddington ratio, winds may form from the innermost accretion region \citep{chartas02,pounds03,reeves03}. Different mechanisms were proposed to explain the formation of the winds \citep[e.g.][]{king03,mizumoto21}.  Associated Fe~K emissions from these flows due to line scattering or recombination in the outflow have been discussed to explain the P-Cygni line profiles in the data\citep[e.g.][]{done07}. Such an emission feature may contribute to part of the Fe~K emission observed in AGN suggested by \citet{parker22} where simulations based on a hybrid model of both razor-thin discs \citep{dauser10} and winds \citep{sim08} are discussed. In \citet{sim08}, the cone-shaped optically thick wind model has a finite thickness too with the inner boundary in a simple linear correlation with $r$ (see Fig.1 in \citet{sim08}). However, \src\ does not show significant evidence of disc winds as in other AGN \citep{tombesi10,parker22}, including either blueshifted Fe~\textsc{xxv}--\textsc{xxvi} \citep{parker16} or blueshifted narrow emissions in the middle energy band \citep{jiang19}. Detailed soft X-ray spectral analysis of \chandra\ grating data of this object also suggests no significant evidence of fast winds forming from the inner region of the disc \citep{lee01}, at least along the line of sight.}   
 
\subsection{The Properties of the Inner Accretion Disc in MCG-06-30-15}

\red{In this section, we discuss other uncertainties concerning the properties of the accretion disc itself in \src\ which we have not taken into account.}

\red{First, the returning radiation of the inner accretion disc has not been considered in our model but may be important in \src. X-ray emission from the innermost region of the accretion disc is subject to the strong gravitational light bending in the strong gravitational field close to the BH. A significant fraction of disc emission may be returned to the disc to be reflected/reprocessed for a second time or multiple times. Such radiation is often referred to as returning radiation.} 

\red{Returning radiation was studied by \citet{cunningham76} and calculated for the disc thermal emission in the X-ray band of stellar-mass BH X-ray binaries. The effects of returning radiation may degenerate with mass accretion rate in the continuum-fitting method \citep{li05}, however, play an important role in polarisation measurements of disc thermal emission \citep{schnittman09}.} 

\red{Similar effects of returning radiation are expected for the non-thermal reflected emission of the disc. \citet{wilkins20} found that the number fraction of returned photons is expected to be the highest ($\approx39\%$) at $a_{*}=0.998$ and the lowest ($6\%$) at  $a_{*}=0$ based on ray-tracing calculations. Both the thin disc model and the razor-thin disc model suggest a high BH spin of $a_{*}>0.87$ in \src. So, at least $\approx20\%$ of the reflected photons of the disc may return to the disc in \src\ \citep{wilkins20}.}

\red{Moreover, we consider a single-zone model with a constant density of $n_{\rm e}=10^{15}$\,cm$^{-3}$ for the disc reflection spectrum in the rest frame of the disc. Recent reflection modelling suggests a higher disc density is required to explain the X-ray data of many AGN and X-ray binaries \citep{ross07,garcia16}. At a high disc density, the soft X-ray band of the disc reflection spectrum shows a blackbody-like emission due to stronger free-free absorption in the reflection slab on the disc. The effects of high disc density are confined in the soft X-ray band. The previous assumption of $n_{\rm e}=10^{15}$\,cm$^{-3}$ is appropriate for massive supermassive BHs with a high Eddington ratio \citep[e.g.][]{jiang18d}. On the contrary, stellar-mass BHs require a very high disc density of $10^{20}-10^{21}$\,cm$^{-3}$ \citep[e.g.][]{tomsick18,jiang19,jiang20}. Mixed results have been found for narrow-line Seyfert 1 galaxies  \citep[e.g.][]{mallick18,jiang20b}, which are believed to host a smaller supermassive BH like the one in \src. Various results found for the discs of low-mass supermassive BHs might be due to their different accretion rates \citep{jiang19b}.}

\red{In the high-density disc reflection model, the soft excess emission commonly seen in AGN is explained as part of disc reflection. The smooth shape of the soft excess emission often requires extreme relativistic effects in the reflection spectrum, e.g. resulting from a high BH spin. \citet{jiang19} compared the spin measurements obtained by different reflection models for the broadband spectra of AGN and found that they were all consistent. However, different spin measurements were found in a few cases where previous work only applied disc reflection models to the Fe~K emission band \citep[e.g. Ton~S180,][]{walton13,parker18,jiang19} and other alternative models were used for the soft excess \citep[e.g.][]{petrucci18}. Due to the ambiguous origin of the soft excess emission, we ignore the soft X-ray band of our data and focus on the prominent features of the disc reflection spectrum of \src, the broad Fe~K emission and the Compton hump. Moreover, the soft X-ray band of \src\ is known to show a mixture of complex dust extinction and warm absorption \citep{lee01}. Nevertheless, we are able to reproduce the similar BH spin measurement using only >3\,keV data when considering the razor-thin disc model as before (see Section\,\ref{m0}).}

\red{One final missing piece in the calculation of rest-frame disc reflection spectra is the radial profile of disc properties, e.g. ionisation. The ionisation parameter is defined as $F/n_{\rm e}$ where $F$ is the illuminating flux that reaches the disc. One may consider the distribution of disc ionisation to be caused by the different illuminating flux in different regions of the disc \citep[e.g.][]{svoboda12}. Such a model considers a constant density $n_{\rm e}$ across the disc. Spectral simulations suggest that such an ionisation distribution may increase the inferred coronal size obtained by the single-zone model \citep{kammoun19}. However, both $F$ and $n_{\rm e}$ are expected to change with radius \citep{shakura73,jiang20}. Therefore, efforts have also been made to calculate radial profiles of disc densities: some consider a simple, phenomenological power law for the radial profile of disc ionisation \citep{askar21} while some consider the same solutions for $P_{\rm rad}$-dominated regions of the standard thin disc in \citet{shakura73} \citep{ingram19}.} 

\red{The calculations of multi-zone disc reflection models are very difficult because 1) the exact density profile of the disc is uncertain. The power of the corona may play an important role in the density profile \citep{svensson94}. Observationally, \citet{haardt93} found that a significant fraction of the disc power may be transferred to the corona, which was also supported by the inferred disc densities of AGN \citep{jiang20b}; 2) detailed ray-tracing is required to calculate the illuminating flux $F$ in different regions of the disc. To do so, a certain geometry of the coronal region, which introduces additional systematic uncertainties as discussed in Section \ref{olet}, has to be considered in the calculation \citep{ballantyne17}; 3) the existence of returning radiation will lead to a steeper radial profile of disc ionisation due to additional illuminating flux from self irradiation \citep{wilkins20}.}

\section{Concluding remarks} \label{discuss}

In this work, we apply the thin disc model \texttt{relxilllpth\_nk} with finite thickness to the hard X-ray data of \src. The mass accretion rate of the disc in \src\ is uncertain because of Galactic absorption. We, therefore, consider a grid of mass accretion rates in the model ranging from $\dot{m}=0$ to 30\%. The lamppost geometry is assumed for the corona of \src. The \texttt{relxilllpth\_nk} calculates consistent disc emissivity profiles during the fit.

We first consider a model with the reflection fraction as a free parameter. By doing so, we obtain consistent BH spin measurements as in previous work, although a slightly higher BH spin and a more compact coronal region are inferred by a higher-$\dot{m}$ model. After taking the disc thickness into consideration, we conclude that the BH spin of \src\ is between 0.87-0.99 (90\% confidence range).

For the first time, we also consider a thin disc model with the reflection fraction consistently calculated for the lamppost geometry. A higher reflection fraction is expected in a thicker disc than in a razor-thin disc. But the difference between the two models is usually less than 1\%. Only in some extreme cases, the difference can reach 10\% at most. 

We argue that the observed values of reflection fraction in \src\ are lower than the theoretical values of this parameter in the lamppost geometry. This explains why significant worse fits are achieved when the lamppost values of $f_{\rm refl}$ are used instead of being treated as a free parameter. Among fits with a linked $f_{\rm refl}$ parameter, the razor-thin disc model provides a better fit to the data than other models with finite thickness. After considering the lamppost value of $f_{\rm refl}$, the razor-thin disc model offers a slightly worse fit than when having a free $f_{\rm refl}$ parameter by $\Delta\chi^{2}=10$.

The overestimation of the reflection fraction parameter in the model suggests that modifications to the over-simplified lamppost geometry are needed for \src\ in future especially when the thickness of the disc is considered. For instance, an outflowing or jet-like corona may exist in \src\ \citep[e.g.][]{wilkins14,gonzalez17}. When the corona is moving away from the BH along the spinning axis, the coronal emission is beamed and the reflection fraction of the disc can be significantly reduced.

\section*{Acknowledgements}

This paper was written during the worldwide COVID-19 pandemic in 2020--2022. We acknowledge the hard work of all the health care workers around the world. We would not be able to finish this paper without their protection. \red{We acknowledge constructive discussion with Michael Parker and Andrew Fabian.} J.J. acknowledges support from the Leverhulme Early Career Fellowship, the Isaac Newton Trust and St Edmund's College, University of Cambridge. The research is supported in part by Grants F-FA-2021-432 and MRB-2021-527 of the Uzbekistan Ministry for Innovative Development.

\section*{Data Availability}

All the data can be downloaded from the HEASARC website at https://heasarc.gsfc.nasa.gov. The thin disc reflection model used in this work will be available \red{at 
https://github.com/ABHModels}.






\bibliographystyle{mnras}
\bibliography{ugc} 



\appendix

\begin{figure*}
    \includegraphics[width=7cm]{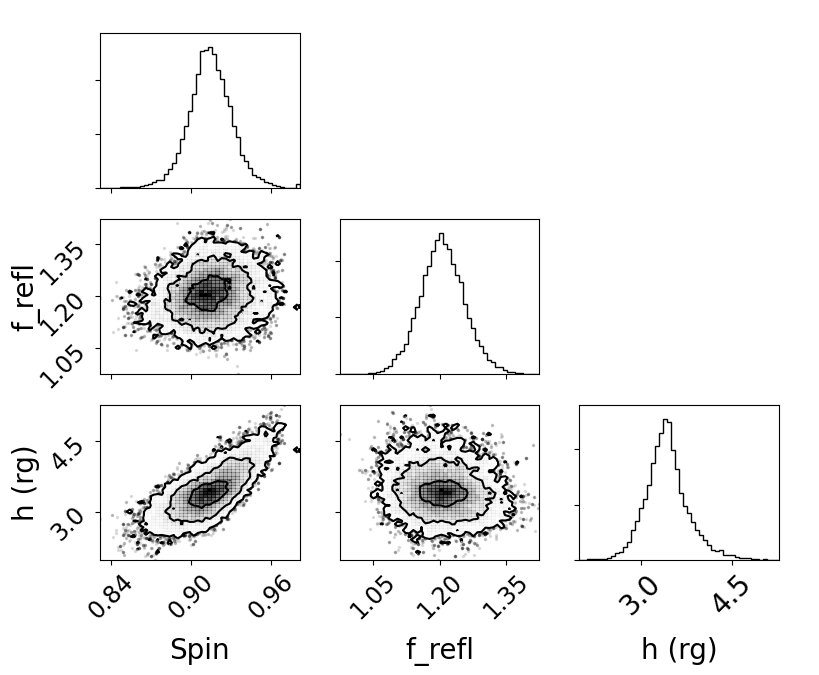}
    \includegraphics[width=7cm]{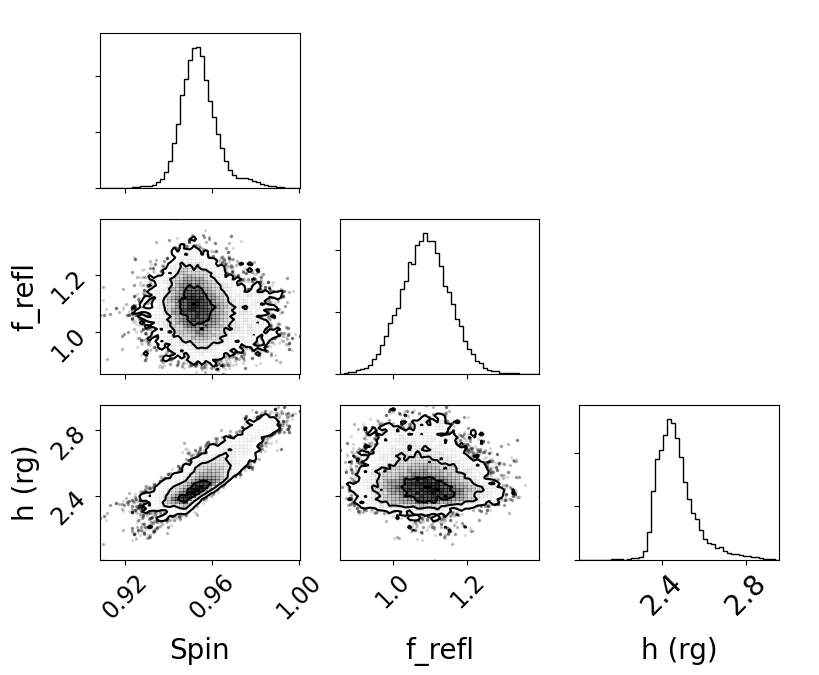}
    \includegraphics[width=7cm]{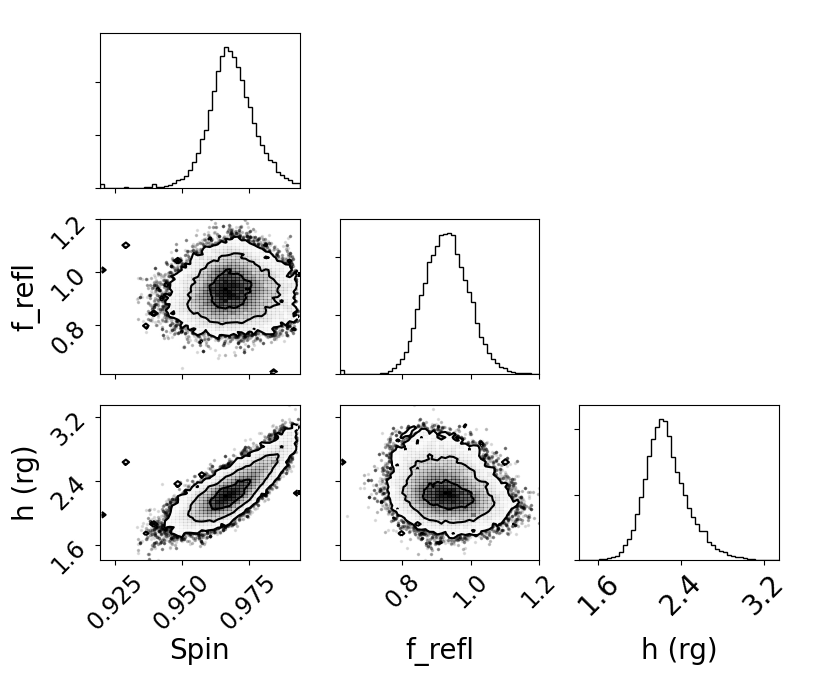}
    \includegraphics[width=7cm]{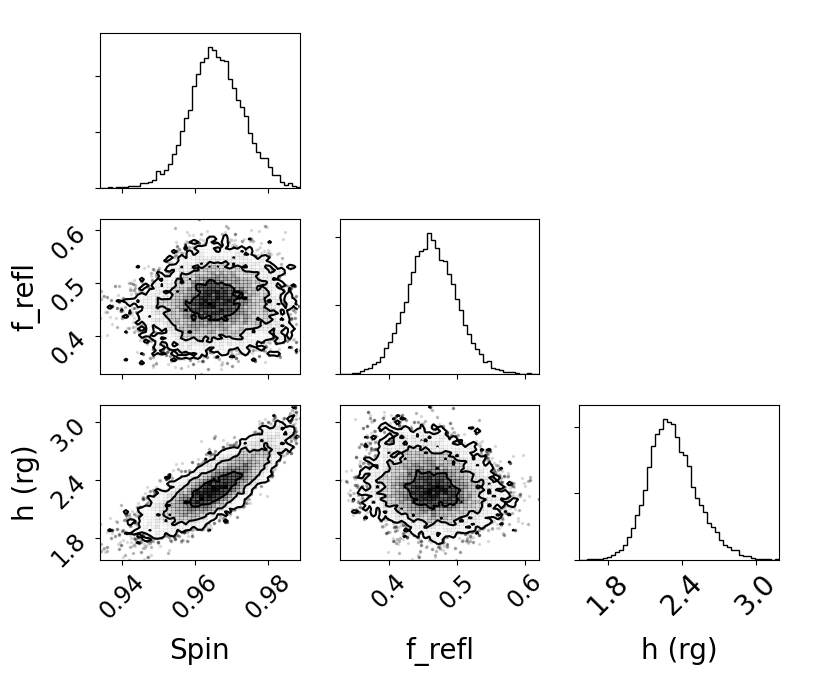}
    \caption{Output distributions for the MCMC analysis of $\dot{m}$=5\% (top left), 10\% (top right), 20\%\ (bottom left) and 30\% (bottom right) for \src.}
    \label{pic_mn0}
\end{figure*}


\bsp	
\label{lastpage}
\end{document}